\documentclass[11pt,a4paper]{article}
\usepackage{jheppub}
\usepackage{subfigure}
\usepackage{hyperref}
\hypersetup{
	colorlinks=true,
	linkcolor=cyan,
	urlcolor=blue,
	citecolor=red,
}

\title{Critical Dynamics in Holographic First-Order Phase Transition}

\author[a]{Qian Chen,}

\author[b]{Yuxuan Liu,}

\author[a,c]{Yu Tian,}

\author[d,e]{Bin Wang,}

\author[f]{Cheng-Yong Zhang}

\author[g]{and Hongbao Zhang}

\affiliation[a]{School of Physics, University of Chinese Academy of Sciences, Beijing 100049, China}

\affiliation[b]{Kavli Institute for Theoretical Sciences (KITS), University of Chinese Academy of Sciences, Beijing 100190, China}

\affiliation[c]{Institute of Theoretical Physics, Chinese Academy of Sciences, Beijing 100190, China}

\affiliation[d]{Center for Gravitation and Cosmology, College of Physical Science and Technology, Yangzhou University, Yangzhou 225009, China}

\affiliation[e]{Shanghai Frontier Science Center for Gravitational Wave Physics, Shanghai Jiao Tong University, Shanghai 200240, China}

\affiliation[f]{Department of Physics and Siyuan Laboratory, Jinan University, Guangzhou 510632, China}

\affiliation[g]{Department of Physics, Beijing Normal University, Beijing 100875, China}

\emailAdd{chenqian192@mails.ucas.ac.cn}

\emailAdd{liuyuxuan@ucas.ac.cn}

\emailAdd{ytian@ucas.ac.cn}

\emailAdd{wang\_b@sjtu.edu.cn}

\emailAdd{zhangcy@email.jnu.edu.cn}

\emailAdd{hongbaozhang@bnu.edu.cn}

\abstract{We study the critical phenomena of the dynamical transition from a metastable state to a stable state in the model of first-order phase transition via two different triggering mechanisms. Three universal stages during the fully nonlinear evolution are extracted. On the one side, by perturbing the scalar source, an isolated seed nucleus is injected into an initial homogeneous state in the supercooled region.
For critical parameters of the seed nucleus, the real-time dynamics reveal that the system will converge to a critically unstable state.
For supercritical parameters, the system exhibits a phase separation, while for subcritical parameters falls back to homogeneous.
The shape independence of the seed nucleus is also investigated, which implies that the critical phenomena are universal.
On the other side, we propose a novel mechanism to render the critical phenomena via a collision of two gravitational shock waves on the dual geometries.
Specifying a collision velocity, the initial system can be also attracted to a critically unstable state.
Aside from these dynamical constructions, we also quantitatively analyze the critical nucleus preventing the system from reaching the final phase separation. We find the depth of the critical nucleus increases almost linearly with the temperature, which implies that the hotter the supercooled state is, the harder for it to trigger phase separation.}


\arxivnumber{}

\begin{document}
	
\maketitle

%=======================================================================
\section{Introduction}\label{sec:In}
First-order phase transition is ubiquitous in nature, and draws a wealth of research, but formulating a theoretical framework is still challenging.
Recently, holography \cite{Maldacena:1997re,Gubser:1998bc,Witten:1998qj,Witten:1998zw} 
provides a new perspective for studying the real-time dynamics of first-order phase transition.
A series of holographic models were constructed in \cite{Gubser:2008ny} to study the equation of state of quantum chromodynamics. 
Specifying different scalar potentials, these models can exhibit either a crossover, first-order, or second-order phase transition.
In the original study of the holographic first-order phase transition, spinodal instability was revealed \cite{Janik:2015iry,Janik:2016btb}, which is similar to Gregory-Laflamme instability \cite{Gregory:1993vy,Gregory:1994bj}. 
The final state of time evolution was found to have an inhomogeneous configuration at the phase transition temperature \cite{Janik:2017ykj,Bellantuono:2019wbn,Attems:2020qkg}, and this dynamical process can be well approximated by hydrodynamics \cite{Attems:2017ezz,Attems:2019yqn}.
Based on these final phase-separated states, the collision dynamics of two adjacent phase domains were studied in Ref. \cite{Bea:2021ieq}.
Taking the finite-size effect into account \cite{Bea:2020ees}, the instability of the intermediate inhomogeneous states was investigated in finite-volume systems. 
Furthermore, generalizing the planar inhomogeneity to circular inhomogeneity, the authors in \cite{Bea:2022mfb} demonstrated that critical bubbles exist when the surface tension balances the pressure difference.

However, the far-from-equilibrium properties of metastable states in the model of first-order phase transition should be further investigated.
On the one side, a dynamical transition occurred when a sufficiently large quench is applied to a metastable state \cite{Li:2020ayr}.
On the other side, from the bulk gravity perspective, the dynamical transition from a metastable state to a stable state is essentially the transition between homogeneous and inhomogeneous black holes. 
Thus, we expect a critical behavior in this dynamical process, according to the critical gravitational collapse \cite{Liebling:1996dx,Choptuik:1996yg,Brady:1997fj,Bizon:1998kq,Choptuik:1992jv,Abrahams:1993wa,Evans:1994pj,Koike:1995jm,Gundlach:1995kd,Garfinkle:1998va,Choptuik:2004ha,Gundlach:2007gc}.

In general, there are two types of critical phenomena in gravitational collapse. % according to the symmetry of the critical solution.
In the type I critical {phenomenon, the time independence of the critical solution renders that the system remains near the critical solution for a period of time satisfying} the relation $\tau\propto -\frac{1}{\lambda}\text{ln}|p-p_{*}|$, where $\lambda$ is the eigenvalue of the \emph{single} unstable eigenmodes of the critical solution.
In the type II critical {phenomenon, the scale independence of the critical solution gives} rise to {a} power-law scaling of the black hole mass at the threshold $M\propto (p-p_{*})^{\frac{1}{\lambda}}$.
Each of these {symmetries} exists in a continuous and a discrete version.

In specific, the type I critical phenomenon in the bald/scalarize black hole transition is recently realized via a nonlinear accretion of the scalar field \cite{Zhang:2021nnn,Zhang:2022cmu,Liu:2022fxy}. Inspired by this thought-provoking setup, in this paper, we would like to investigate similar critical phenomena of the dynamical transition from a metastable state to a stable state in the holographic model of the first-order phase transition, and analyze universal critical behaviors in this dynamical transition.

The plan of the paper is as follows:
In section \ref{sec:Hs}, we introduce a holographic first-order phase transition model and show the phase structure.
In section \ref{sec:Cp}, we present the critical dynamics in the dynamical transition from an initial supercooled state to a final phase-separated state. First, various forms of seed nuclei are introduced to produce the critical behavior of the system and the corresponding critical exponent is fitted.
Second, we impose a collision velocity on the initial system to excite gravitational shock waves on the dual geometries \cite{Attems:2018gou,Chesler:2010bi,Casalderrey-Solana:2013aba,Casalderrey-Solana:2013sxa,Chesler:2015wra,Chesler:2015bba,Chesler:2015lsa,Chesler:2016ceu,Attems:2016tby,Attems:2017zam,Attems:2016ugt}, which also leads to the convergence of the system to a critically unstable state.
Next, we elaborate directly on the properties of the critical nucleus in this critical state. 
In section \ref{sec:Di}, we conclude the paper by a summary and an outlook.

%=======================================================================

%=======================================================================
\section{Holographic setup}\label{sec:Hs}
In this section, we will introduce a holographic model of the first-order
phase transition, and then present the phase structure in thermodynamics.

\subsection{Modeling the first-order phase transition}
First, we consider the Einstein's gravity coupled to a real scalar field with the action
\begin{equation}
	S=\frac{1}{2\kappa^{2}_4}\int d^{4}x\sqrt{-g}\left[ R-\frac{1}{2}(\nabla\phi)^{2}-V(\phi)\right]+S_{\partial{M}},
	\label{eq:action}
\end{equation}
where $\kappa^{2}_4=8\pi G_4$ is set to $1$ for convenience and the boundary term $S_{\partial M}$ includes the Gibbons-Hawking action \cite{Gibbons:1976ue} and the holographic counterterm \cite{Bianchi:2001kw,Elvang:2016tzz}.
The form of the potential function is taken the same as in \cite{Janik:2017ykj}
\begin{equation}
	V(\phi)=-6{\rm cosh}(\frac{\phi}{\sqrt{3}})-\frac{\phi^{4}}{5}
	\label{eq:potential}
\end{equation}
for a holographic first-order phase transition.
With the AdS radius $L=1$ and conformal dimension of the scalar operator $\Delta=2$, the near-boundary expansion of the scalar field can be obtained as
\begin{equation}
	\phi=\phi_{1}r^{-1}+\phi_{2}r^{-2}+o(r^{-3}),
	\label{eq:asy_scalar}
\end{equation}
where the source $\phi_{1}$ is chosen to be $1$, while the response $\phi_{2}$ {remains} undetermined by the {near-boundary} analysis, {and its value can only be obtained after solving the bulk.}

{In this model, we attempt to} study the nonlinear dynamics of an initial homogeneous state under a large inhomogeneous disturbance only in the $x$ direction. The corresponding process can be solved by the characteristic formulation \cite{Chesler:2013lia}, and hence we adopt the following ansatz
\begin{equation}
	ds^{2}=\Sigma^{2}(Gdx^{2}+G^{-1}dy^{2})+2dt(dr-Adt-Fdx),
	\label{eq:EF}
\end{equation}
where {the translational invariance in the $y$ direction is preserved, and} all metric {components} are functions of $(t,x,r)$. 
The form of the metric ansatz is {invariant under} the shift transformation in the radial coordinate $r\rightarrow \overline{r}=r+\lambda$, which {is used to fix the apparent horizon at} $r=1$ during evolution.
The asymptotic near-boundary behavior of metric {components can be expressed as}
\begin{equation}
	\begin{aligned}
		G&\sim 1+g_{3}r^{-3}+o\left(r^{-4}\right),\\
		\Sigma&\sim r+\lambda - \frac{1}{8}r^{-1}-\frac{1}{24}\left(\lambda+4\phi_{2}\right)r^{-2}+o\left(r^{-3}\right),\\
		F&\sim -\partial_{x}\lambda+f_{1}r^{-1}+o\left(r^{-2}\right),\\
		A&\sim \frac{1}{2}\left(r+\lambda\right)^{2}-\frac{1}{8}-\partial_{t}\lambda+a_{1}r^{-1}+o\left(r^{-2}\right).\\
	\end{aligned}
	\label{eq:asy_metirc}
\end{equation}

{The corresponding scalar operator $\left\langle O\right\rangle$ and energy-momentum tensor $\left\langle T_{ij}\right\rangle$ of the dual theory can be read off as}
\begin{equation}
	\begin{aligned}
		\left\langle O\right\rangle =&\frac{1}{2}\left(\lambda+\phi_{2}\right),\\
		\left\langle T_{ij}\right\rangle
		=&\begin{pmatrix}
			-2a_{1}-\left\langle O\right\rangle&-\dfrac{3}{2}f_{1}&0\\
			-\dfrac{3}{2}f_{1}&-a_{1}+\dfrac{3}{2}g_{3}&0\\
			0&0&-a_{1}-\dfrac{3}{2}g_{3}\\
		\end{pmatrix},
	\end{aligned}
	\label{eq:one-point}
\end{equation}
{which satisfies the Ward identities as}
\begin{equation}
	\left\langle T^{i}_{i}\right\rangle=\left\langle O\right\rangle,\qquad \nabla^{i}\left\langle T_{ij}\right\rangle=0,
\end{equation}
{with the source $\phi_{1}=1$.} 

\subsection{Phase diagrams and instabilities}
In order to find the phase structure, {first we need to} solve the field equations to get the static homogeneous solutions.
In this case, { we find $G=1$, $F=0$, and the remaining fields $\Sigma,A,\phi$ can be solved} by the Newton-Raphson iteration algorithm.
Since the geometry has been determined, the temperature and entropy density of the system can easily be extracted respectively {as}
\begin{equation}
T=\frac{\partial_{r}A|_{r=r_{H}}}{2\pi},\qquad s=2\pi\Sigma^{2}(r_{H}),
\end{equation}
where $r_{H}$ represents the location of the event horizon.
The free energy density satisfies the following thermodynamic relation
\begin{equation}
	\mathcal{F}=\left\langle T_{tt}\right\rangle-sT
\end{equation}
where $\left\langle T_{tt}\right\rangle $ is the $tt$ component of the Brown-York tensor (\ref{eq:one-point}), which stands for the energy density of the boundary theory.

%%%%%%%%%%%%%%%%%%
\begin{figure}
	\centering
		\subfigure[]{\includegraphics[width=.50\linewidth]{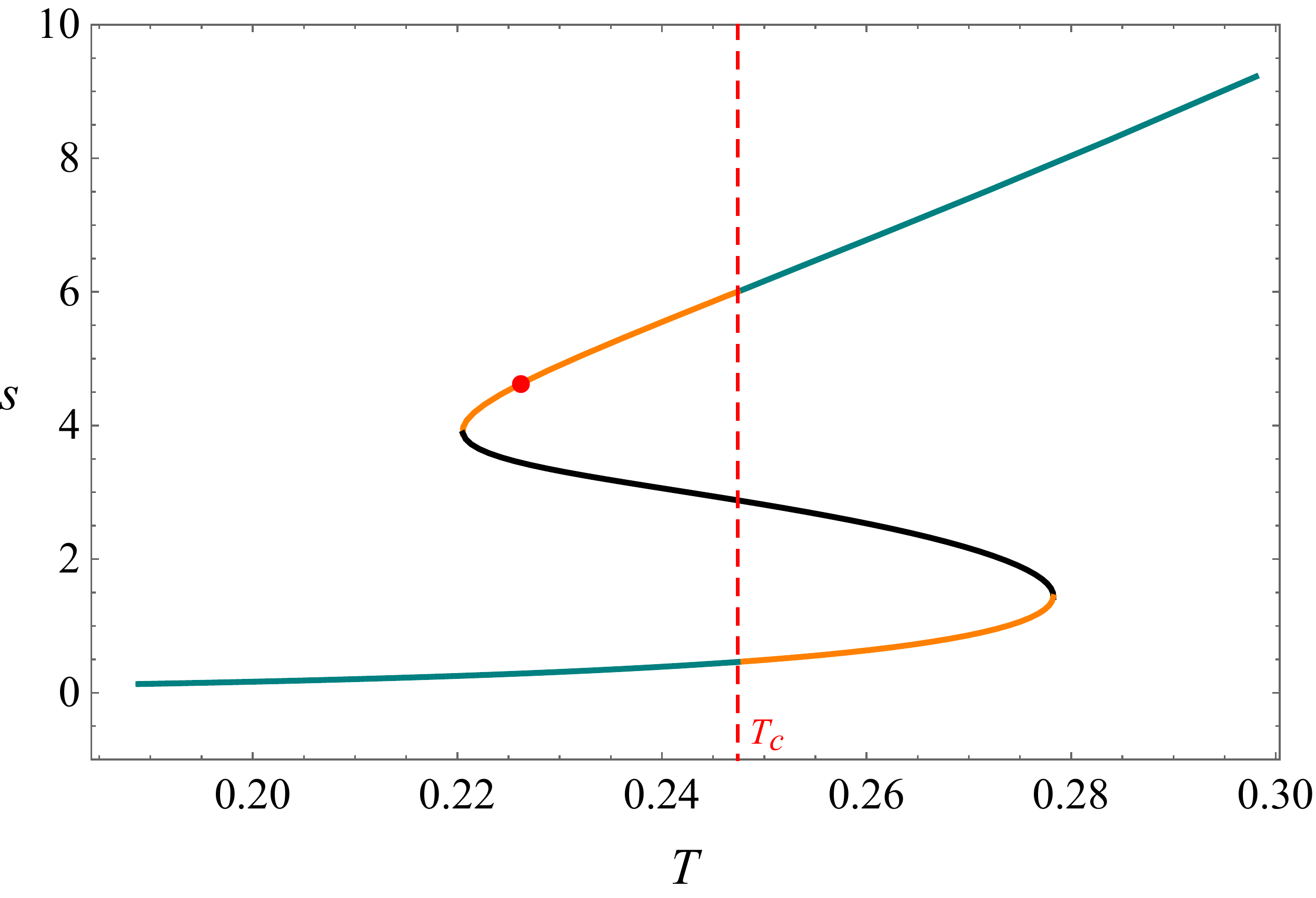}\label{fig:EOS_F}}
		\subfigure[]{\includegraphics[width=.48\linewidth]{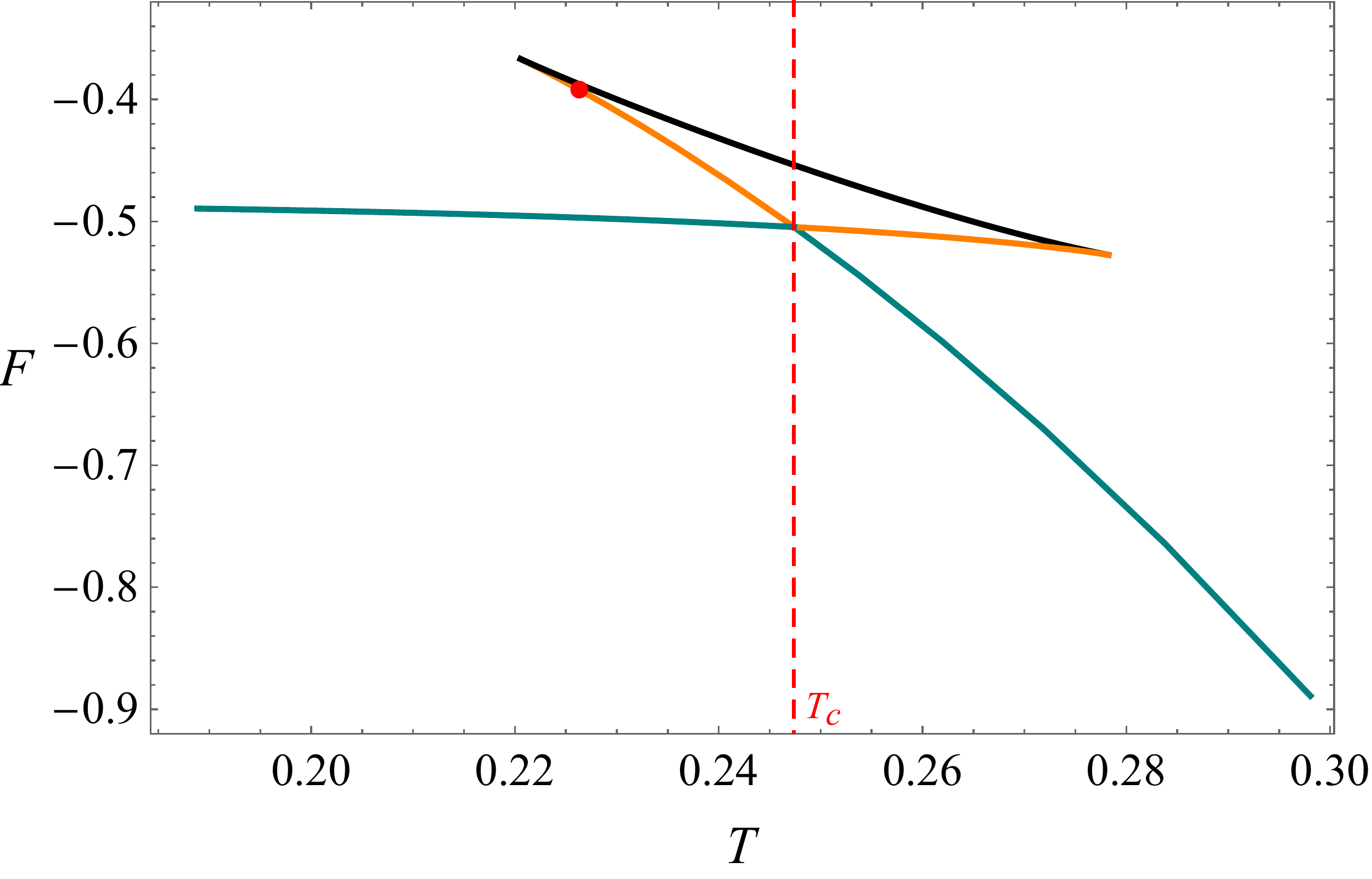}
		\label{fig:EOS_S}}
		\caption{(a): The free energy density and (b): entropy density of the  of static homogeneous solution versus the temperature.
			The vertical red dotted line represents the phase transition temperature $T_{c}=0.24739\pm 0.00001$.
			The red point is the initial state of the system under consideration.}
		\label{fig:EOS}
\end{figure}
%%%%%%%%%%%%%%%%%

\begin{figure}
	\begin{center}
		\includegraphics[height=.35\textheight]{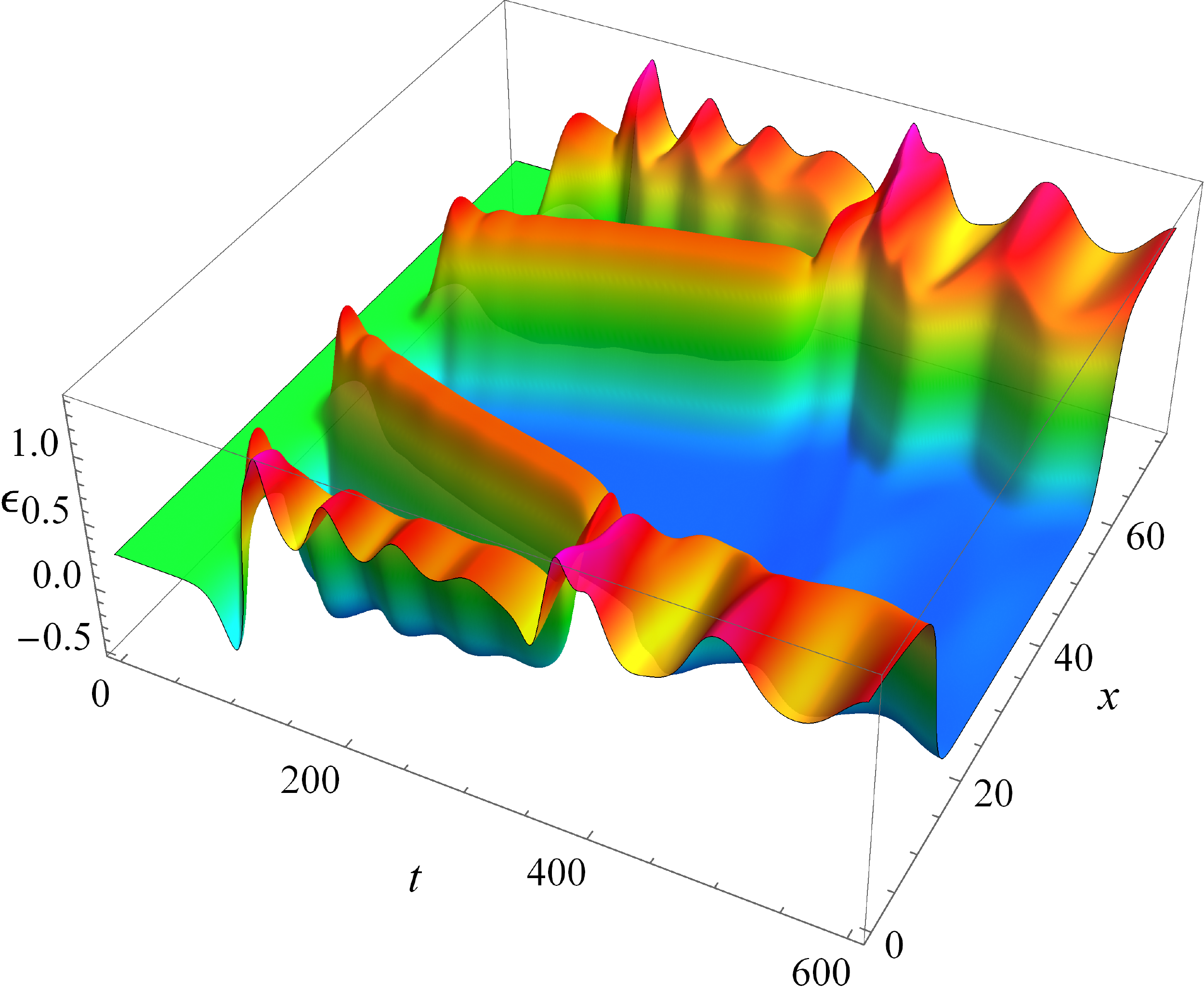}
		\caption{In the spinodal region, any arbitrarily small perturbation will drive the system to a final phase-separated state \cite{Janik:2017ykj} (see also \cite{Bellantuono:2019wbn,Attems:2020qkg,Bea:2020ees}). For instance, in this plot, the initial state with $\{T,s\}=\{0.25,2.77\}$ is perturbed by a Gaussian-like perturbation in the form of $\delta\phi=-0.1r^{-4}(r-1)^{2}\text{exp}\left[-10\text{cos}^{2}\left(\frac{x}{24}\right)\right]$, and the subsequently nonlinear evolution of the energy density possesses a final phase separation.}
		\label{fig:spinodal}
	\end{center}
\end{figure}

The static homogeneous solutions of the model are shown in Fig.~\ref{fig:EOS}.
Fig.~\ref{fig:EOS_F} and Fig.~\ref{fig:EOS_S} are the functions of free energy and entropy density with respect to temperature, respectively.
The states can be divided into three types as highlighted in different colors according to the stability of the system.
First, the states in cyan are stable, 
and two discontinuous cyan branches at $T=T_{c}$ in Fig.~\ref{fig:EOS_S} reveal the existence of first-order phase transition on account of the discontinuous first-derivative of the free energy -- Fig.~\ref{fig:EOS_F}. \footnote{It is worth noting that the dynamical transition we explored in this paper is not the aforementioned thermodynamic phase transition, but is triggered by a critical disturbance, which we will introduce in the next section.}
Second, the states in black are both locally thermodynamically and dynamically unstable. The reason for the former is the negative specific heat of the system, while the latter is in agreement with the Gubser-Mitra conjecture \cite{Gubser:2000ec,Gubser:2000mm}, which is confirmed by the analysis based on linear response theory \cite{Janik:2015iry} and nonlinear dynamical evolution \cite{Janik:2017ykj}.
Due to the dynamical instability, the {states in this region will always evolve to some phase-separated states shown in Fig. \ref{fig:spinodal}, with the free energy in both phases being equal}.
Third, the remaining states in orange are metastable, which means they are linearly stable but nonlinearly unstable, since in these cases, the phase separation will still arise from a large inhomogeneous disturbance shown in Fig. \ref{fig:real_time_energy_density}.

%%%%%%%%%%%%%%%%%%
%%%%%%%%%%%%%%%%%

In the next section, we will explore the critical phenomena during the dynamical transition from a metastable state to a stable state via two different triggering mechanisms, and then demonstrate the existence of the critical nucleus in the intermmediate unstable state.

%=======================================================================
\section{Critical dynamics}\label{sec:Cp}
\begin{figure}
	\begin{center}
		\includegraphics[height=.26\textheight]{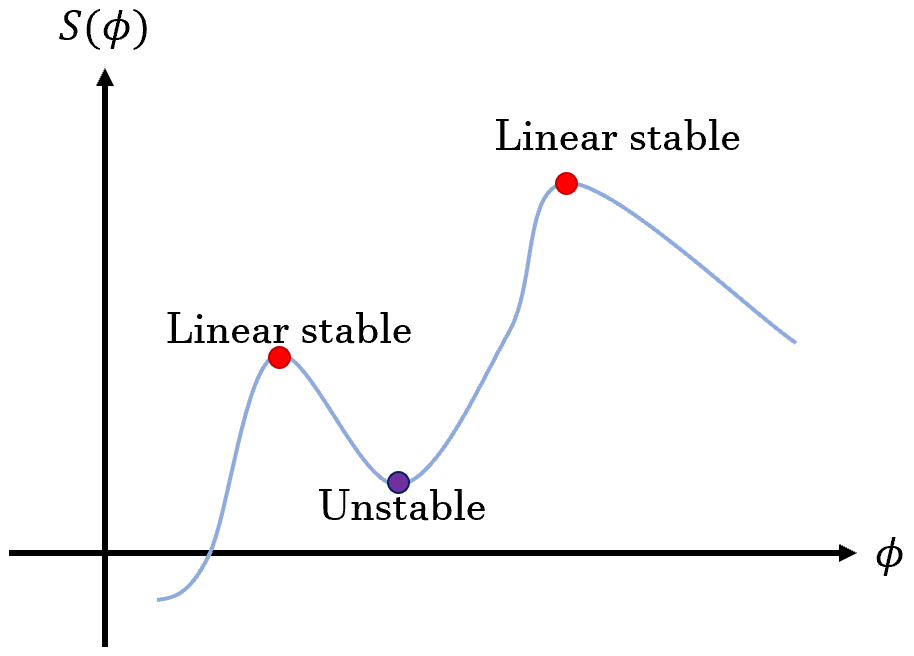}
		\caption{A sketch of entropy landscape with thermodynamic entropy being a function of scalar field configurations. The curve in light blue represents solutions to equations of motion. In specific, two red dots denote static solutions with linear stability, while the purple dot denotes a static solution that is unstable.}
		\label{fig:landscape}
	\end{center}
\end{figure}

Critical phenomena widely exist during dynamical evolution in the presence of a dynamical transition when the initial state suffers from a nonlinear instability.
From the viewpoint of the entropy landscape (Fig.~\ref{fig:landscape}), two linearly stable states with local maximal entropy must be separated by an unstable state with an entropy at a saddle point. If the unstable state possesses only one unstable mode, then two linearly stable states can be connected by one single-parameter curve family \cite{Laine:2016hma,Gould:2021ccf,Bea:2022mfb}. Specifically, in our case, the unstable state can be approached via tuning the single parameter of a quench. \footnote{That is to say, the type I critical behavior occurs in this process.} In fact, it is argued in \cite{Li:2020ayr} that such an unstable state acts as the lowest dynamical barrier for the quench to trigger the transition, though from the viewpoint of the free energy landscape in the canonical ensemble.

%=======================================================================

The above entropy landscape is just a rough picture to help understand the physics in the critical dynamics here. As has been done in \cite{Li:2020ayr} for the free energy landscape in the canonical ensemble (though in the probe limit), it is expected that a precise description of an entropy landscape should also include at least: 1) An exact definition of the configuration space (the horizontal axis in Fig.~\ref{fig:landscape}), such that the thermodynamic quantities (the entropy here) can be calculated; 2) A form of the dynamical evolution equations in terms of the functional gradient on the landscape as the driving force. It is still unclear whether the form of the driving force for a precise entropy landscape can be fully achieved in the back-reacted, inhomogeneous case.

In this section, we will first present the fully nonlinear evolution introduced by a seed nucleus. Then, we will investigate the similar critical behavior in the collision of two gravitational shock waves in the dual geometries. Finally, we will analyze the properties of the dynamical barriers in detail.

\subsection{Critical behavior induced by seed nuclei}
In the freezing process of a supercooled liquid,
a seed nucleus is required to trigger a dynamical transition. Similarly, we also expect a similar structure in the framework of holographic first-order phase transition.
In the spinodal region, the dynamical transition occurs no matter how small the perturbation. While in the metastable region, there is a dynamical barrier to generate a dynamical transition. Therefore, any metastable state must be quenched at least to a critically unstable state to trigger the dynamical transition.

%%%%%%%%%%%%%%%%%%
\begin{figure}[t]
	\begin{center}
		\subfigure[]{\includegraphics[width=.49\linewidth]{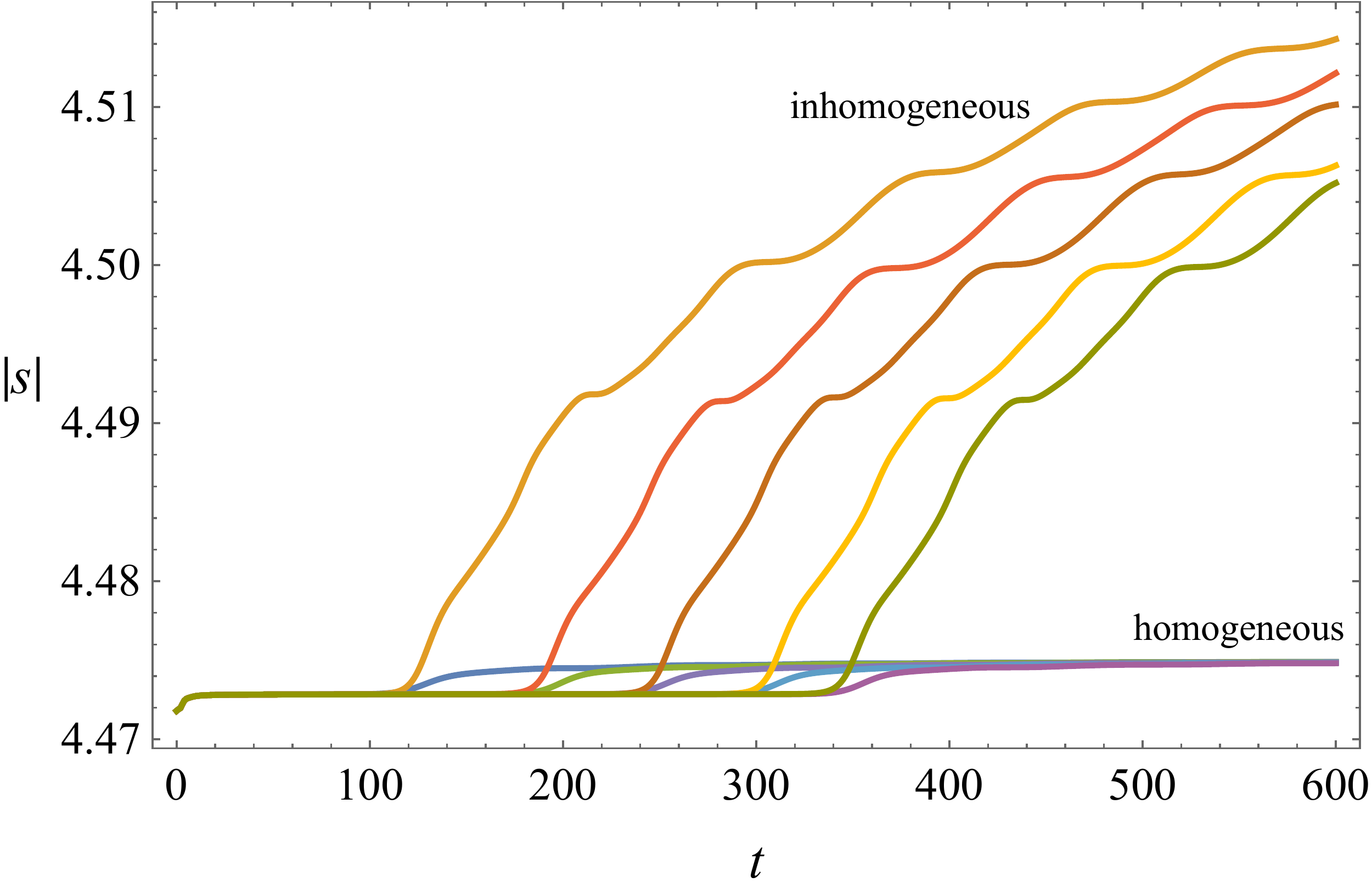}\label{fig:critical_behavior1}}
		\subfigure[]{\includegraphics[width=.49\linewidth]{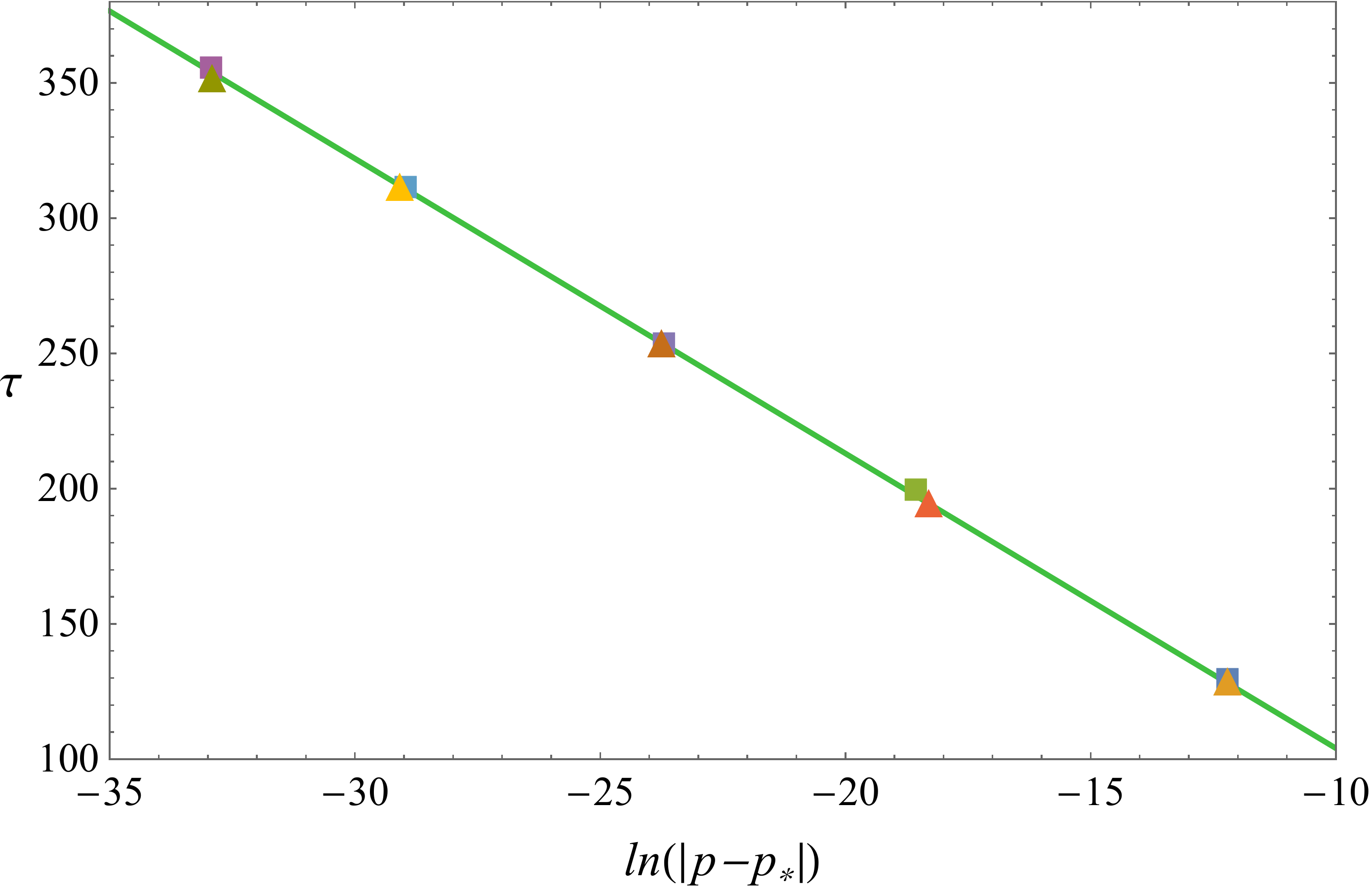}\label{fig:critical_behavior2}}
		\caption{(a): The evolution of the average entropy for various $p$ close to the {threshold} $p_{*}\approx 0.929654988651265$. The time-dependent entropy density is defined as $s=2\pi\Sigma^{2}_{H}$, where $\Sigma_{H}$ represents the value of the metric field $\Sigma$ at the apparent horizon. The average entropy of the system is defined as $|s|=\int sdx/(L_{x}\pi)$. The curves in different colors represent the corresponding dots in (b). Specifically, the triangle and square dots in (b) correspond to the inhomogeneous and homogeneous final states respectively.
		(b): The duration of the system near a critical solution with respect to $\text{ln}|p-p_{*}|$. Note that the calculation of the time $\tau$ actually includes the first stage for convenience. It does not affect the critical exponent since the time for the first stage is the same in all considered cases.}
		\label{fig:critical_behavior}
	\end{center}
\end{figure}
%%%%%%%%%%%%%%%%%

\subsubsection*{Gaussian-like seeds}
{Without loss of generality, we choose the red point in the supercooled region as the initial state -- Fig.~\ref{fig:EOS}}, and the following families of disturbance served as seed nuclei are added to the scalar field
\begin{equation}
	\delta \phi=ar^{-2}{\rm exp}\left[ -w{\rm cot}^{2}(\frac{x}{L_{x}})\right],
	\label{eq:disturbance1}
\end{equation}
where the parameters $a,w$ are the amplitude and width of the Gaussian wave, respectively.
The length of the box is equal to $L_{x}\pi$, where the parameter $L_{x}$ is set to $24$ to minimize the {finite-size} effect on the physical results.
For initial data parameterized by $p$, which could be $a$ or $w$ here, there is a critical $p_{*}$ that just trigger the dynamical transition from a metastable state to a stable state.
Without loss of generality, {the strategy we use is to} fix the width $w=50$ and {then, vary} the amplitude $a$ to {find the critical phenomena}.

\begin{figure}
	\begin{center}
		\subfigure[]{\includegraphics[width=.49\linewidth]{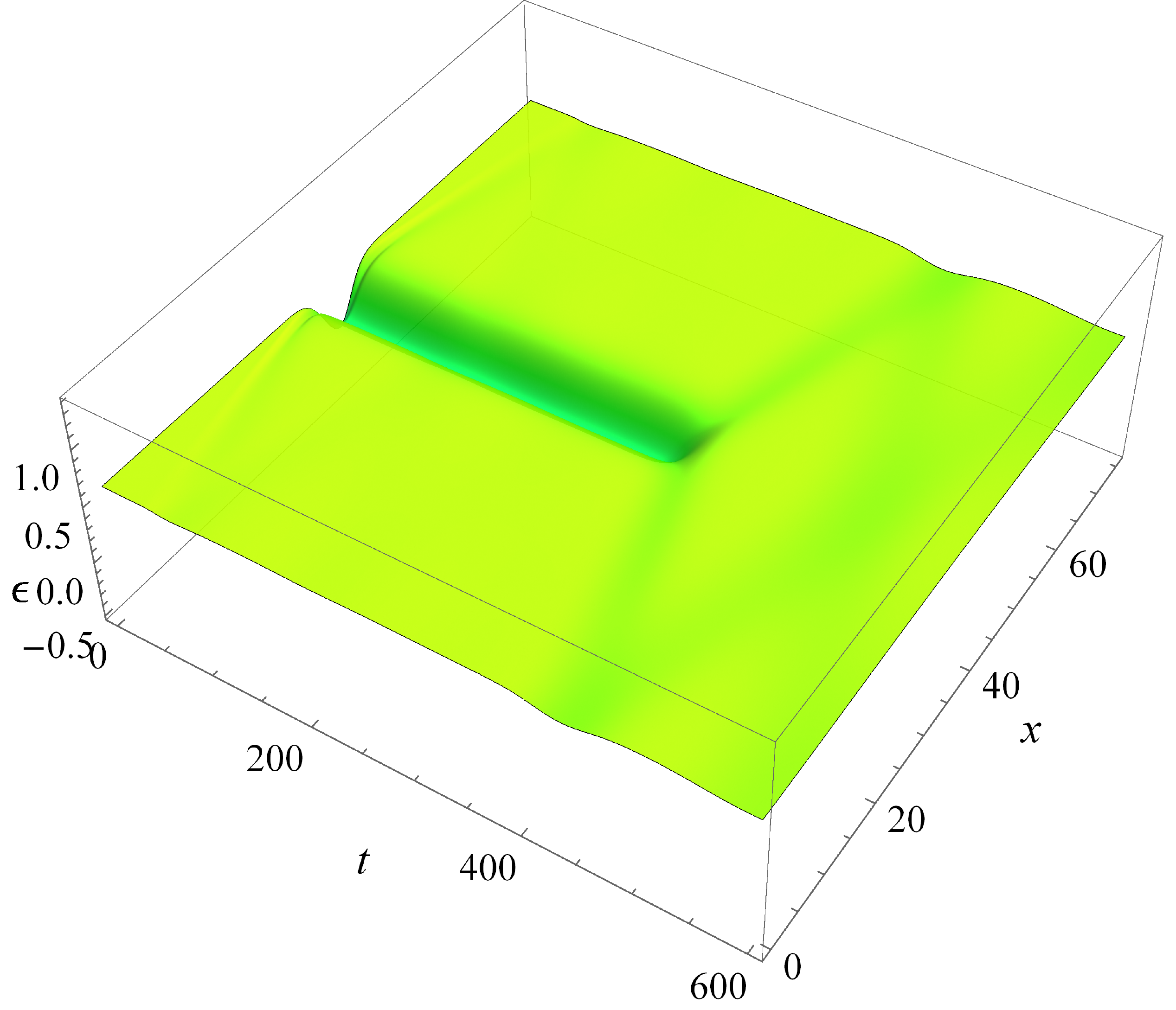}\label{fig:real_time_energy_density1}}
		\subfigure[]{\includegraphics[width=.49\linewidth]{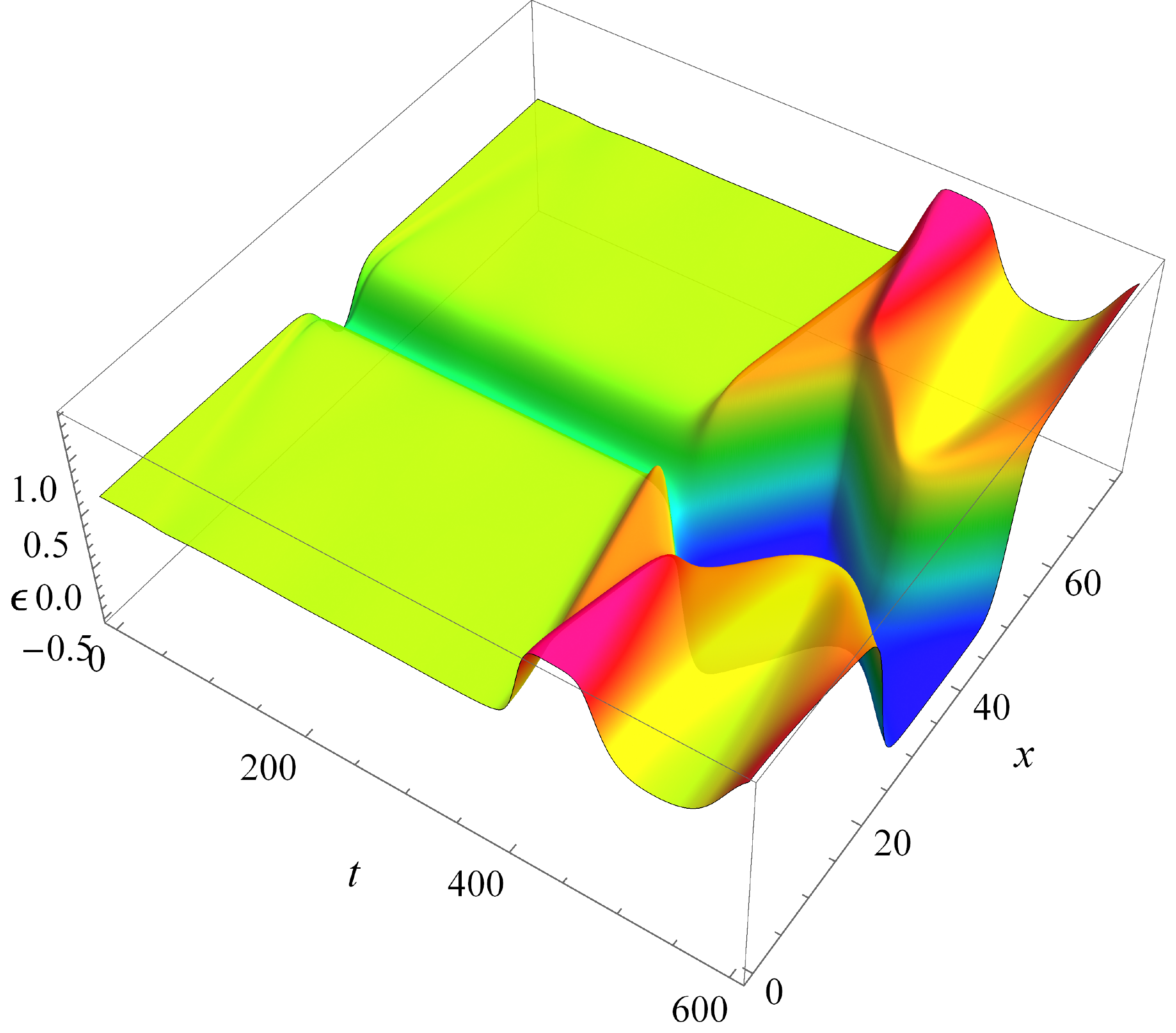}\label{fig:real_time_energy_density2}}
		\caption{The energy density as a function of time with the parameter $p$ is closest to the threshold $p_{*}$ in the cases (a): $p<p_{*}$ and (b): $p>p_{*}$.}
		\label{fig:real_time_energy_density}
	\end{center}
\end{figure}

The time dependence of the average entropy of the system is shown in Fig. \ref{fig:critical_behavior1}. When parameter $p$ continuously increases from the subcritical to supercritical cases, the corresponding entropy of the final state undergoes a discontinuous jump. Therefore, it is essentially a first-order dynamical transition.
The overall evolution process can be divided into three stages.
In the first stage, all initial data characterized by the parameter $p$ close to a threshold $p_{*}$ converge rapidly to the vicinity of a critically unstable state, like the effect of an attractor.
In the second stage, the solutions enter the linear region of the critical solution, which can be well approximated by
\begin{equation}
	\phi(t,X)\approx\phi_{*}(X)+\left(p-p_{*}\right)e^{\eta t}\delta\phi(X)+\text{decaying modes},
	\label{eq:linear_region}
\end{equation}
where $X=(x,r)$.
The static configuration $\phi_{*}(X)$ {stands} for the critical solution, and $\delta\phi(X)$ is  the \emph{only} unstable eigenmode associated with the eigenvalue $\eta$.
Since the unstable modes are dominated at the second stage, the period of
time the system near the critical solution, satisfies the relation 
\begin{equation}
    \tau\propto-\gamma \text{ln}(|p-p_{*}|),\label{eq:exponent}
\end{equation}
where the critical exponent $\gamma=\eta^{-1}=10.9$ for both subcritical and supercritical cases.
This is shown in Fig.~\ref{fig:critical_behavior2}, analogous to the cases in type I critical gravitational collapse. 
At late times of this stage, the unstable eigenmodes push the system exponentially away from the {critical} solution when the simulation time $t>\tau$.
In the final stage, the critical states evolve into homogeneous states for subcritical parameters $p<p_{*}$, while into inhomogeneous states for supercritical parameters $p>p_{*}$. The final inhomogeneous states consist of two stable phases with the same value of free energy, and these different phases are connected by domain walls -- Fig.~\ref{fig:real_time_energy_density}. 
Due to the nonlinear instability of the initial states, the evolution possesses a critical phenomenon in the intermediate stage, which must be triggered by a sufficiently large disturbance. Therefore, this phenomenon is obviously different from that found in \cite{Janik:2017ykj}, where the initial states suffer from linear instability, and arbitrarily small perturbation will lead to final phase separation.
Moreover, the constant temperature of the final inhomogeneous state is equal to the phase transition temperature $T_{c}$, which means that this two-phase-coexisting system is in thermal equilibrium.

%%%%%%%%%%%%%%%%%%
%%%%%%%%%%%%%%%%%

To clearly show {the critical behavior within} the three stages, the temporal and spatial dependence of the energy density {are} displayed in the Fig.~\ref{fig:real_time_energy_density}. 
Fig.~\ref{fig:real_time_energy_density1} and ~\ref{fig:real_time_energy_density2} correspond to the {case} $p<p_{*}$ and $p>p_{*}$, respectively. 
There is a critical inhomogeneous state in the intermediate process -- see also in \cite{Bea:2020ees}, and the initial homogeneous system will cross this critical unstable state and subsequently evolve to a stable inhomogeneous system only when the seed nucleus is sufficiently large. Otherwise, the system will return to the homogeneous state.
In general, it takes more time to reach a stable inhomogeneous state, than a final homogeneous state, due to the smaller value of the imaginary part of the dominant decaying mode, and the difficulty of reaching a stable inhomogeneous state increases with the depth of the critical nucleus (see Sec.~\ref{sec:CriticalSolution} for more discussions), which depicts the energy well deviating from the homogeneous configuration.

%%%%%%%%%%%%%%%%%%
\begin{figure}
	\begin{center}
		\subfigure[]{\includegraphics[width=.49\linewidth]{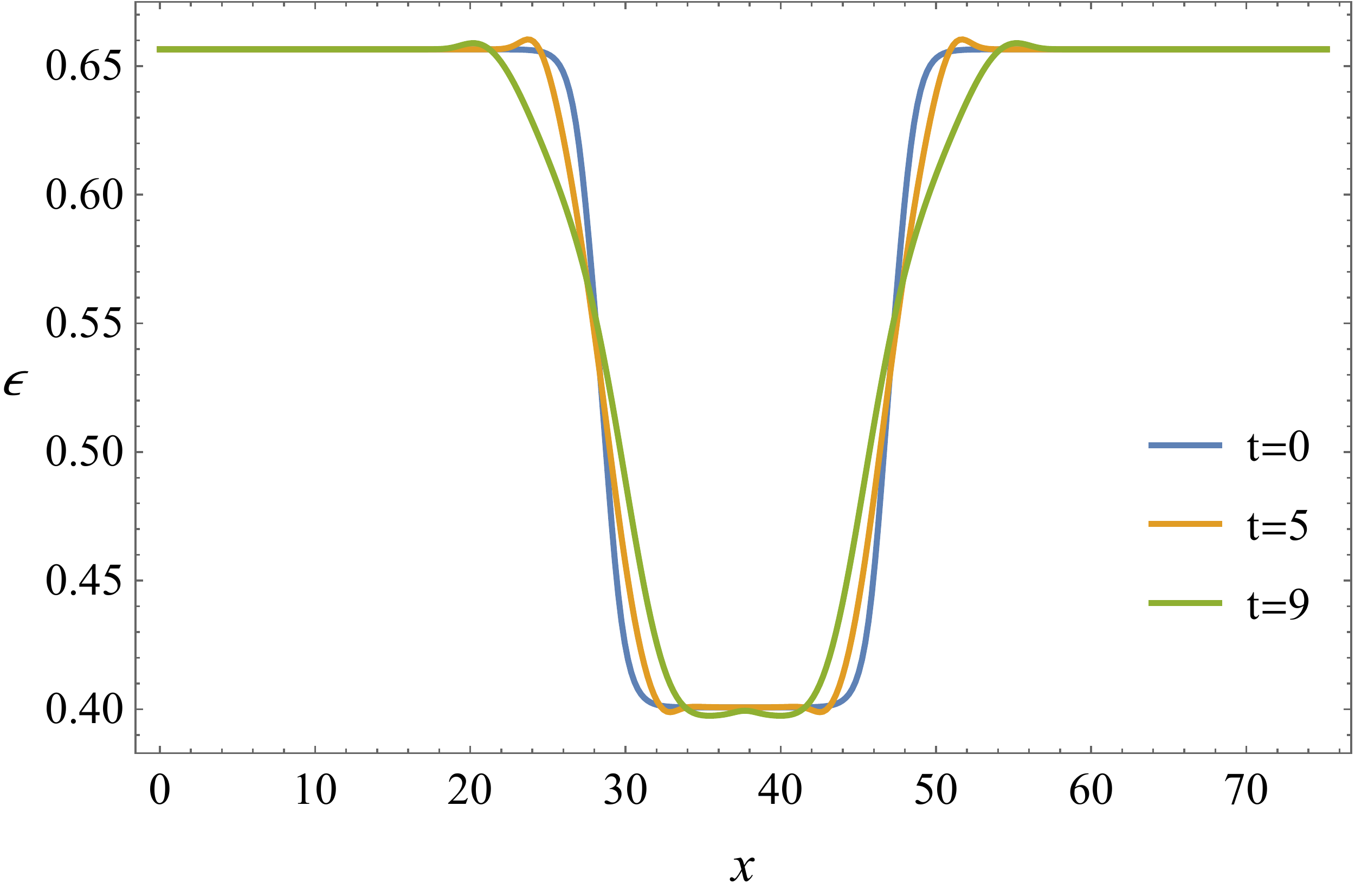}\label{fig:nucleus1}}
		\subfigure[]{\includegraphics[width=.49\linewidth]{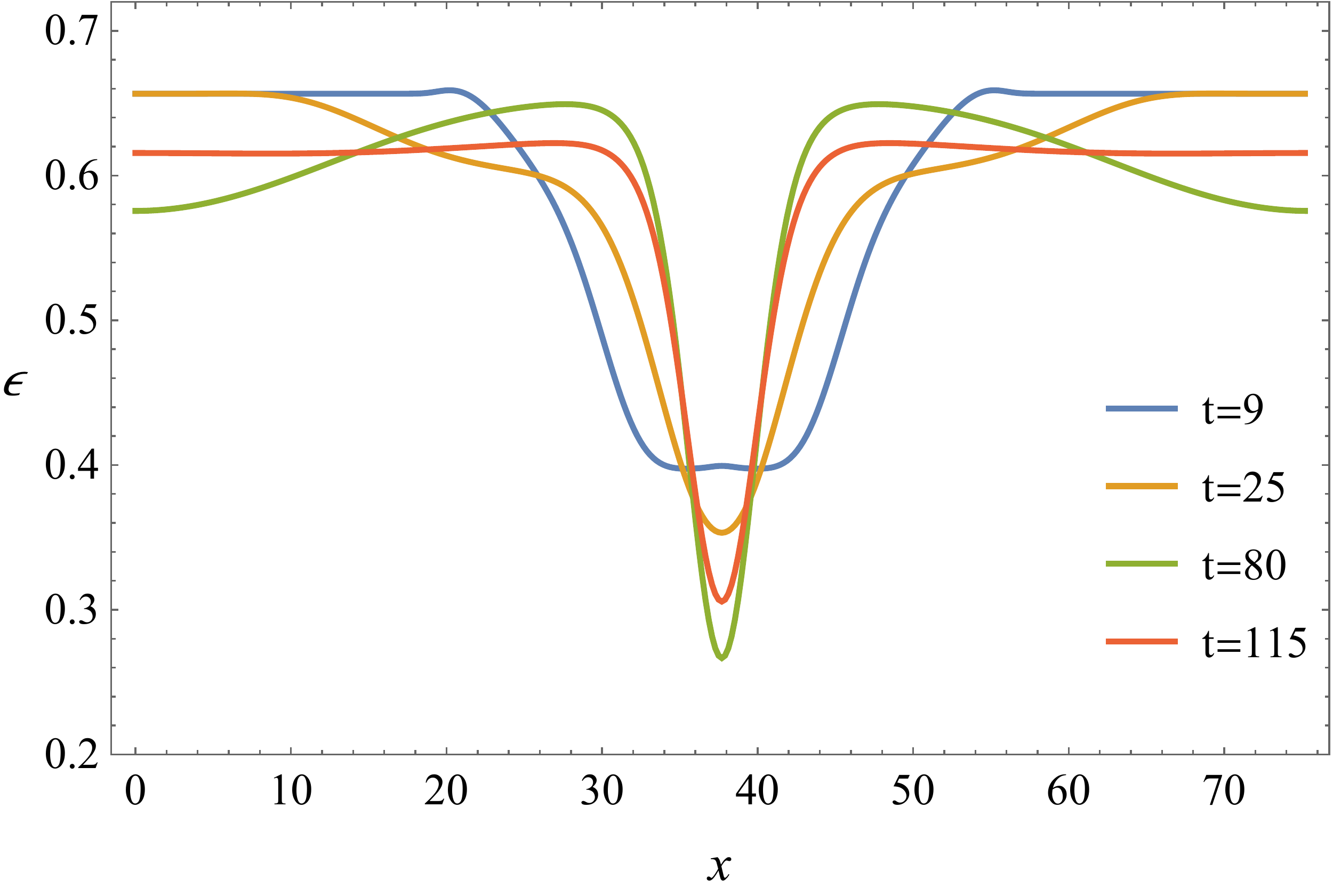}\label{fig:nucleus2}}
		\caption{The energy density at different times in the process where the system converges to the critical solution.  The parameter $\delta$, which characterizes the length of the nucleus, is fixed to $3\pi$, and parameter $\alpha$ is set to $0.8L_{x}$ to keep the maximum numerical error below $10^{-9}$.}
		\label{fig:nucleus}
	\end{center}
\end{figure}
%%%%%%%%%%%%%%%%%

\subsubsection*{General forms of seeds}
The critical phenomenon in these real-time dynamics is universal, and independent of the specific form of seeds, e.g. (\ref{eq:disturbance1}). Taking a more general form of seed nuclei,
\begin{equation}
	\delta \phi=br^{-2}f\left( \frac{x+\delta}{L_{x}}\right)f\left( \frac{\delta-x}{L_{x}}\right),
	\label{eq:disturbance2}
\end{equation}
where the function $f(x)$ is defined as $\frac{1}{2}\left[ 1+\tanh (\alpha\cot x)\right]$, the critical behavior remains qualitatively the same.
This type of seed nuclei exhibits a rectangular shape on the initial state shown in Fig.~\ref{fig:nucleus1}. The critical solution can be obtained by tuning the parameter $b$ and the {corresponding subsequent evolution is shown in Fig.~\ref{fig:nucleus}, where two characteristic stages during evolution, are illustrated in Fig.~\ref{fig:nucleus1} and Fig.~\ref{fig:nucleus2}, respectively.}

In the first stage, the dynamical behavior of the system is independent of the existence of the critical nucleus{, and the} system is propelled by the pressure difference between the inside and outside of the rectangular nucleus imposed by the disturbance (\ref{eq:disturbance2}). 
{At first, the} system exhibits {a} local dynamical instability at the domain walls due to the pressure gradient{, and there} is an energy flow from the {high-energy region} through the domain wall to the {low-energy region}, and the geometry tends to {be homogeneous -- Fig. \ref{fig:nucleus1}}.
{As the process goes on}, the attractive effect of {the critical nucleus becomes significant} when the two energy flows converge in the middle.
In the second stage, the local configuration of the system changes drastically and quickly converges to {a sharp} critical nucleus {-- Fig. \ref{fig:nucleus2}}.
{This process} can be well approximated by (\ref{eq:linear_region}).
Simultaneously, the energy flow decays in space, and the state approaches an unstable critical solution. The final stage is similar to the former case. For subcritical parameters, the system falls back to homogeneous, while for the supercritical parameters, the system evolves to a phase-separated state.

%%%%%%%%%%%%%%%%%%
\begin{figure}
	\begin{center}
		\includegraphics[height=.35\textheight]{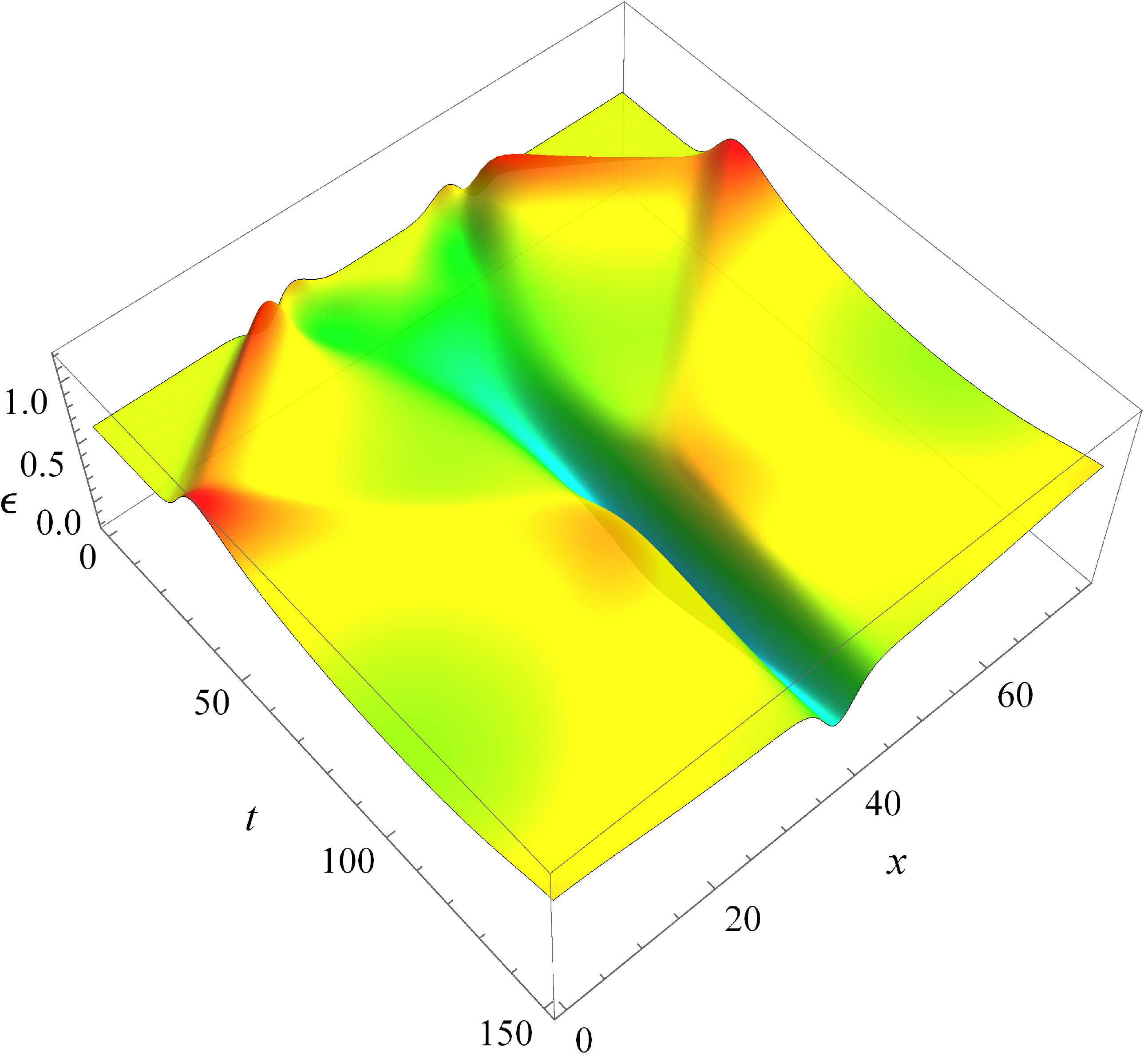}
		\caption{The energy density as a function of time in the collision process of shock waves. Note that in this plot, we have only illustrated the fully nonlinear evolution up to the intermediate critical state.}
		\label{fig:collision}
	\end{center}
\end{figure}
%%%%%%%%%%%%%%%%%

\subsection{Critical behavior induced by collisions}
Holographic collisions were extensively studied in literature \cite{Attems:2018gou,Chesler:2010bi,Casalderrey-Solana:2013aba,Casalderrey-Solana:2013sxa,Chesler:2015wra,Chesler:2015bba,Chesler:2015lsa,Chesler:2016ceu,Attems:2016tby,Attems:2017zam,Attems:2016ugt}, which reveal plenty of intriguing properties of strongly coupled, non-Abelian plasma, especially for the quark-gluon plasma created in the relativistic heavy-ion collisions. In the dual geometries, it can be described by the collision of two gravitational shock waves. In this subsection, we will demonstrate that critical phenomena also exist in this physical scenario, which can be viewed as a novel mechanism for triggering the critical behavior.

\subsubsection*{Gaussian-like shock waves}
In the model of first-order phase transition, the collision of shock waves is realized by applying a collision velocity to a supercooled state. Without loss of generality, we still choose the initial state represented by the red dot in Fig. \ref{fig:EOS}.
In order to obtain two low-energy waves moving towards each other, the momentum density of the boundary theory $\left\langle T_{tx}\right\rangle $ is imposed {by} the following families of disturbance:
\begin{equation}
	\delta\left\langle T_{tx}\right\rangle=-\frac{3}{2}\left[ g(\frac{x-\delta}{L_{x}})-g(\frac{x+\delta}{L_{x}}) \right] ,
	\label{eq:disturbance3}
\end{equation}
where the Gaussian-like function $g(x)=H\text{exp}\left[ -w\text{cot}^{2}(x)\right]$. 
The distance between the centers of the two wave packets is $2\delta$, where $\delta$ is set to $4\pi$.

This kind of excitation provides two head-on colliding wave packets on top of the initial homogeneous state. The intermediate critical state can be approached by fine-tuning
the amplitude $H\to H_{*}$ and the fully nonlinear evolution is illustrated in Fig.~\ref{fig:collision}.

%%%%%%%%%%%%%%%%%%
\begin{figure}
	\begin{center}
		\subfigure[]{\includegraphics[width=.49\linewidth]{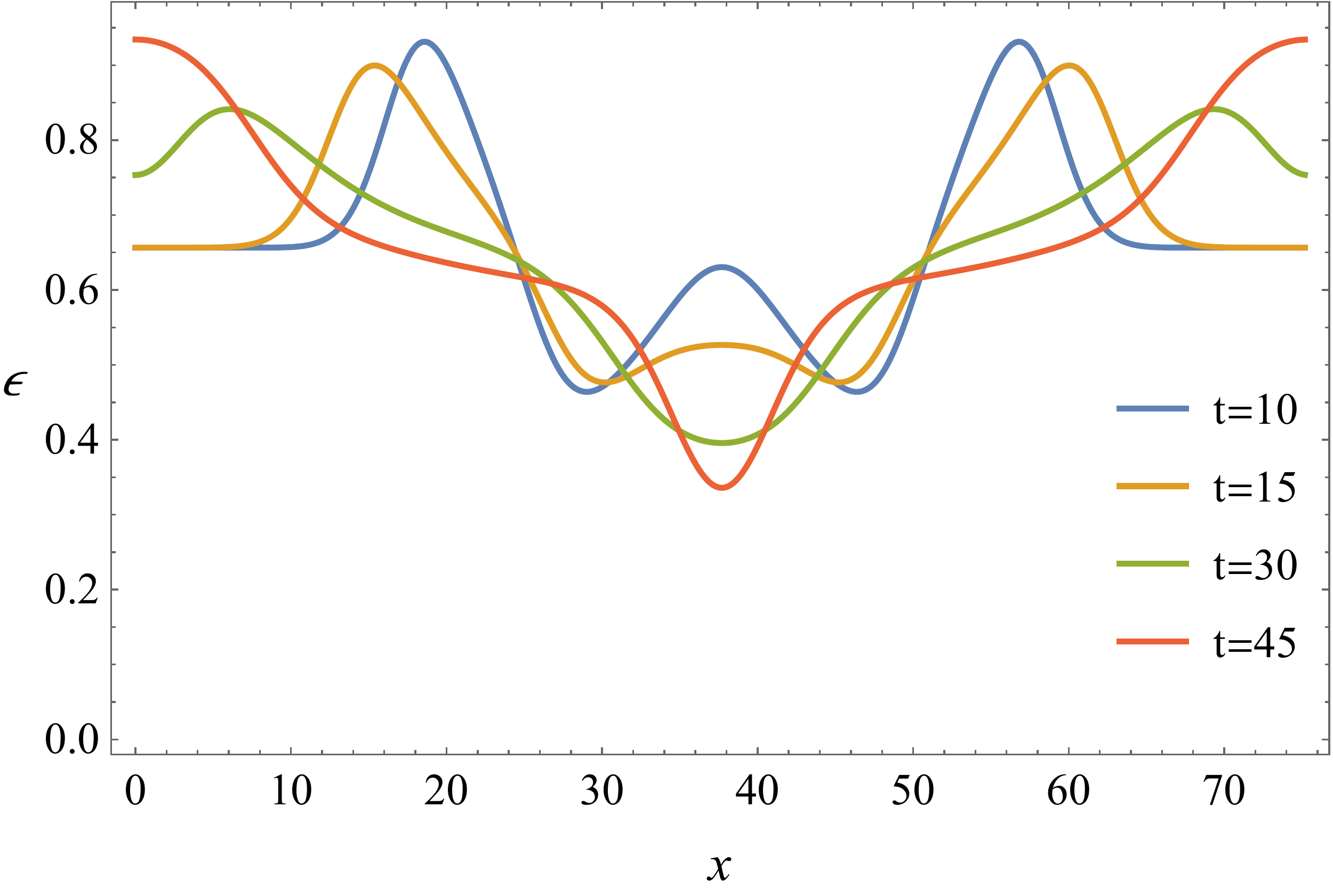}\label{fig:collsion2}}	
		\subfigure[]{\includegraphics[width=.49\linewidth]{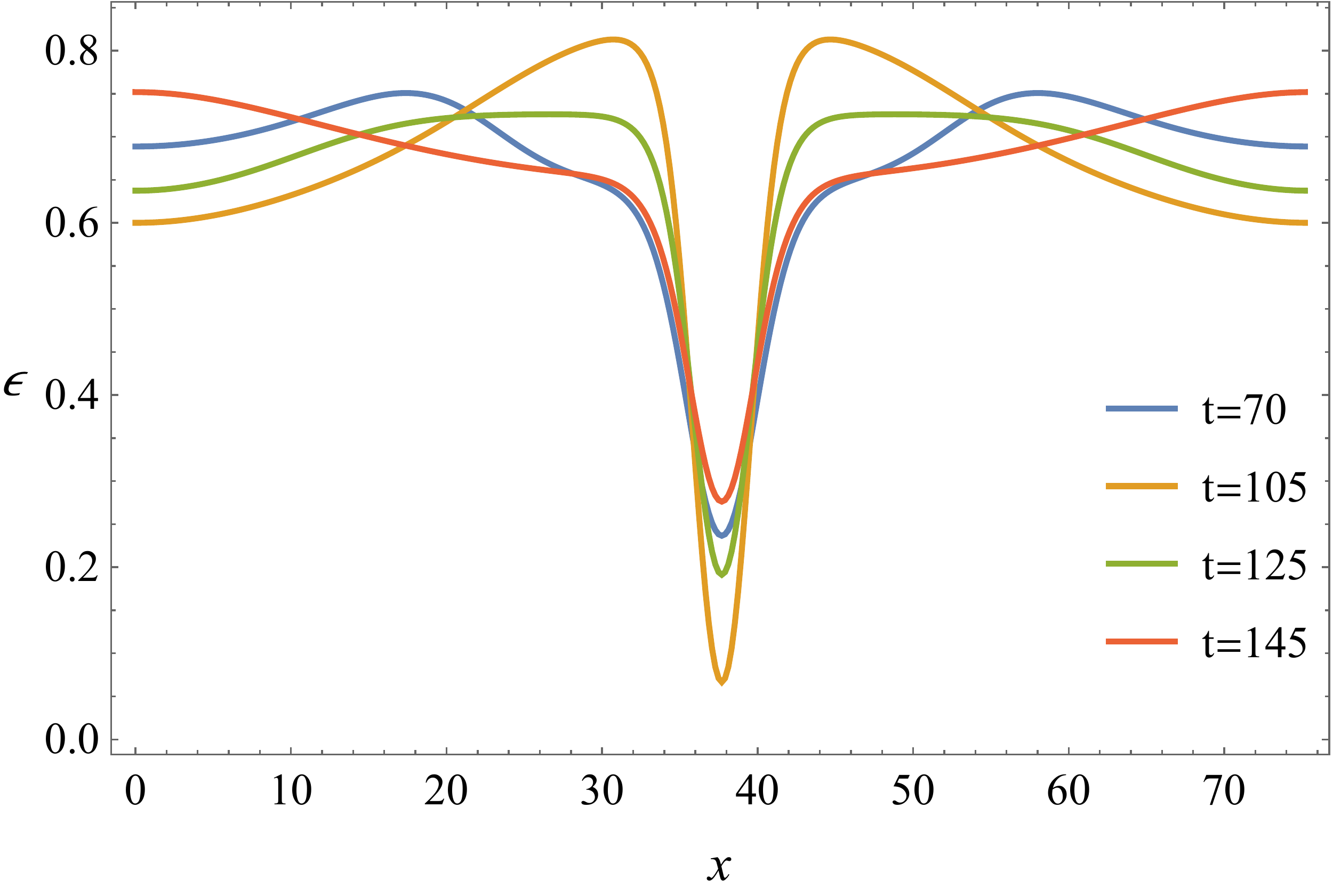}\label{fig:collsion3}}
		\caption{(a): Merger of inward waves and damping of outward waves until $t \approx 30$; The formation of critical nuclei and injection of outward waves around $t \approx 45$. (b) Two damping waves hit the domain wall around $t \approx 105$, and subsequently bounce back.}
	\end{center}
\end{figure}
%%%%%%%%%%%%%%%%%%

Three characteristic stages can also be distilled from the fully nonlinear evolution. By applying the collision velocity (\ref{eq:disturbance3}), the system will be perturbed by two low-energy waves moving inward and two high-energy waves moving outward. At the very beginning of the first stage, the inward waves will collide at the center of the phase domain and then merge together, while the outward waves will damp with propagation. For a nearly critical parameters, the merged waves are finally attracted to the critical nucleus configurations -- Fig.~\ref{fig:collsion2}. Meanwhile, due to the periodic boundary conditions, outward waves will inject from the other boundary. These two damping waves will hit the domain wall and then bounce back -- Fig.~\ref{fig:collsion3}. In the second stage, the solution also persists a similar critical behavior. The unstable modes dominates most of this stage. In the final stage, the system will evolve to a phase-separated state only for supercritical parameters, while to a homogeneous state otherwise.

From the above observations, we find that only the last two stages are similar to the previous cases, but the dynamics of the first stage, namely the stage before the emergence of the critical phenomena, is totally different. That is the main reason we call it a novel triggering mechanism.

Regardless of not only the specific form of the disturbance but also the triggering mechanism, dynamical processes can always be divided into three distinct stages, which illustrates the universality of critical dynamics.

%=======================================================================

\subsection{Unstable states and critical nucleus}\label{sec:CriticalSolution}
The unstable critical states, at which the critical phenomena emerge, are static solutions at the intermediate stage, and can be approximated by fine-tuning parameters to a critical threshold during fully nonlinear evolution. As intuitively illustrated in the previous subsections, the critical nuclei in these unstable states act as an energy well, and depict the difficulty to form a final phase-separated state. Therefore, in this subsection we will further investigate the role played by the critical nucleus in the dynamical transition.

%%%%%%%%%%%%%%%%%%
\begin{figure}
	\begin{center}
		\subfigure[]{\includegraphics[width=.50\linewidth]{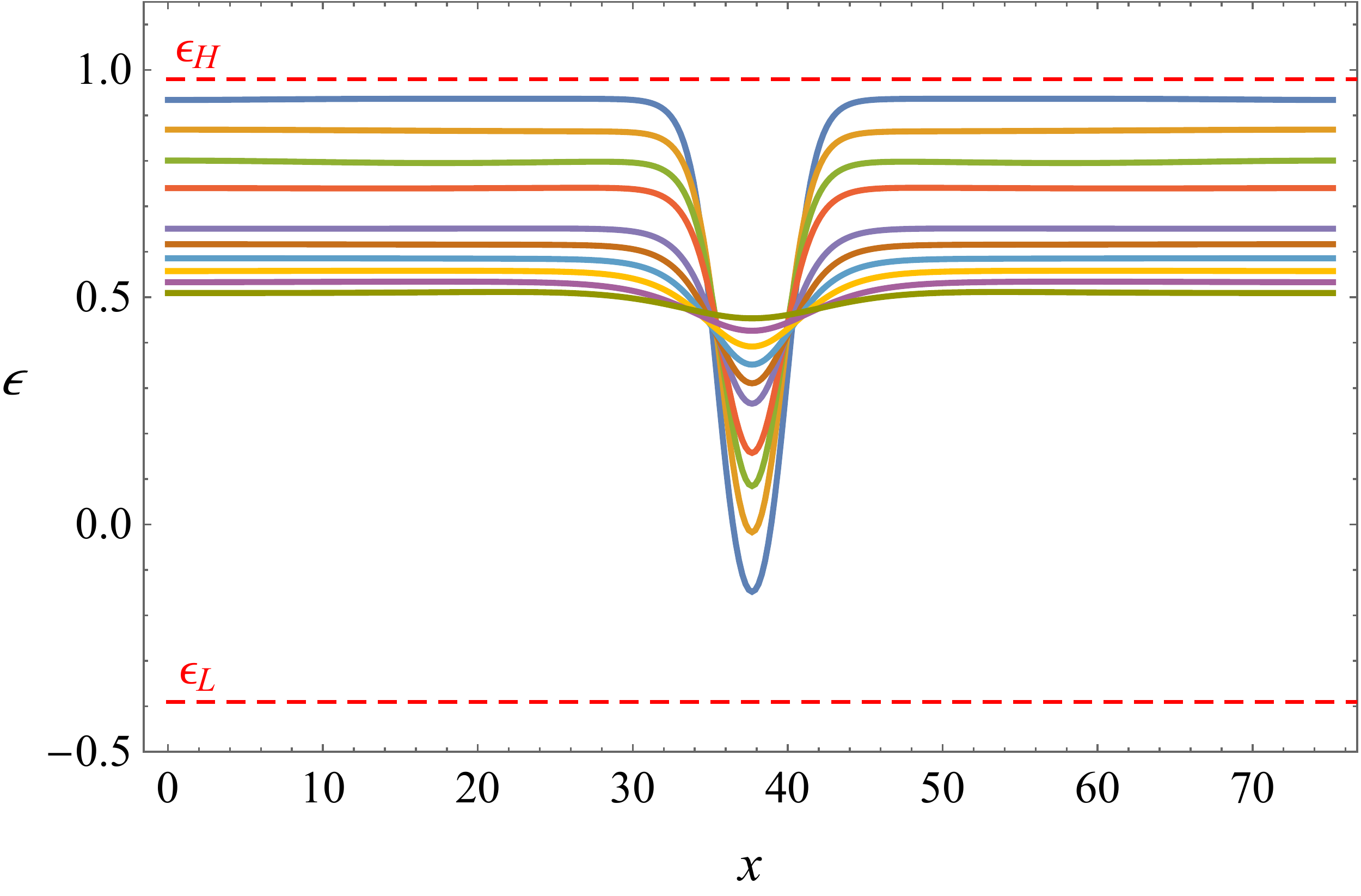}
			\label{fig:shape}}	
		\subfigure[]{\includegraphics[width=.48\linewidth]{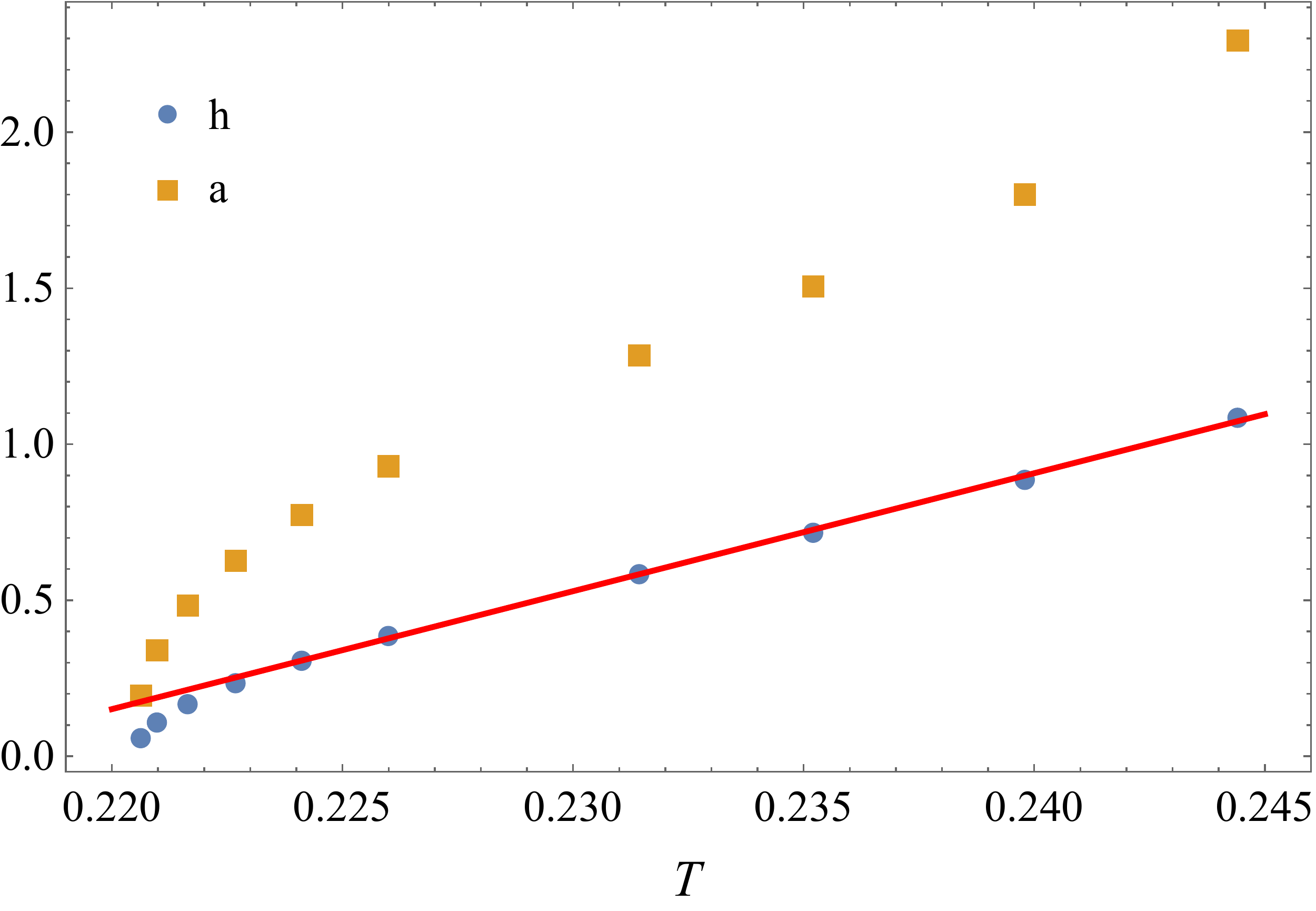}\label{fig:size}}
		\caption{(a): The energy densities of the critical nucleus at different temperatures. The horizontal red dashed lines represent the energy densities at the phase transition temperature in homogeneous solutions. (b): the depth of critical nuclei (blue dots) and the amplitude of Gaussian-like seeds (orange squares) with respect to temperature.}
	\end{center}
\end{figure}
%%%%%%%%%%%%%%%%%%

As shown in Fig.~\ref{fig:shape}, for every intermediate unstable state at different temperature, there is a critical nucleus at the center of the domain. The tendency of the amplitude of seed nuclei required to provide critical phenomena is almost the same as the depth of critical nuclei in the corresponding state -- Fig.~\ref{fig:size}. In this sense, the critical nuclei can be viewed as a good approximation of the energy well to reach a final phase-separated state.
As expected, the depth of critical nuclei increases monotonically with temperature -- Fig.~\ref{fig:size}. Thus, a general tendency is that the colder the supercooled state, the more likely phase separation is to occur. More specifically, for the lower-temperature region in the supercooled domain, the depth of critical nuclei shrinks to zero with a fractional power-law behavior. This is reasonable since the states in this region are close to the spinodal domain -- Fig.~\ref{fig:EOS_S}. At the same time, in most regions, the depth of critical nuclei increases linearly with temperature.

These energy wells can further be quantitatively verified at the experimental level. Due to the nearly constant width of critical nuclei at different temperatures, the amount of energy that needs to be extracted from a critical nucleus can be approximated by its depth -- Fig.~\ref{fig:shape}. That is to say, following the prediction of Fig.~\ref{fig:size}, by specifying the temperature of a system, one is capable to determine the total energy that needs to be extracted from a fixed nucleus, and produce critical phenomena.

In one-dimensional cases, there is no pressure difference at the domain wall, and we quantitatively described the energy well to form a phase separation by illustrating the configurations of critical nuclei at different temperatures. The effect of pressure difference is investigated in the two-dimensional case \cite{Bea:2022mfb} by introducing a circular bubble, which can be compared to experiments, such as ice formation \cite{bai2019probing}.

%=======================================================================
\section{Conclusion}\label{sec:Di}
In this paper, we have investigated the critical dynamics in the holographic model of first-order phase transition triggered by a critical disturbance. Specifically, the fully nonlinear evolution of a supercooled state in the metastable domain is provided by two different mechanisms, which reveal the underlying far-from-equilibrium properties of the system. For the first mechanism, the critical phenomena are triggered by an isolated seed nucleus, while for the second mechanism, the critical phenomena occur because of a collision of two gravitational shock waves on the dual black hole geometries.

In both mechanisms, We can distill three characteristic stages during evolution. In general, the first stage is quite different for the two mechanisms, while the last two stages are almost the same: in the first stage, when injecting an isolated seed nucleus, the supercooled state will be rapidly attracted to an intermediate critical state. Nevertheless, when introducing a holographic collision, the inward shock waves will merge in the middle, acting like a seed, while the damping outward shock waves will periodically bounce back from the critical nucleus. At the end of the day, the system will also be attracted to a critical state at a much larger time scale. In the second stage, for a nearly critical parameter, the system will persist in an intermediate critical state. With the dominance of the unstable mode (\ref{eq:linear_region}), the period of time the system remains on this state satisfies (\ref{eq:exponent}). At late times of this stage, the system starts to move away from the critical state. In the third stage, the system will finally evolve to a phase-separated state for supercritical parameters, or a homogeneous state for subcritical parameters.

Finally, we also analyze the properties of the intermediate critical states. Within these states, the energy density possesses a low-lying land in the middle of the domain, which we call the critical nucleus. During the fully nonlinear evolution, the critical nucleus plays the role of energy well to obtain a final phase separation. To create a critical nucleus, the energy needs to be extracted from a fixed region increases almost linearly with temperature, which is expected to be tested in future experiments.

The widespread presence of critical phenomena manifests its universality in the model of first-order phase transition. Essentially, as long as a sufficiently large disturbance is given to initial metastable states, the system will possess a critically dynamical process, from which three universal stages can be distilled.

The inspiration for the critical phenomena explored in this paper comes from the type I critical gravitational collapse. Therefore, a natural question is whether there is the type II critical phenomena {\cite{Brady:1997fj}} in the models of holographic phase transition.
Since the critical solution in the type II critical phenomena is scale-invariant, there is more interesting physics worth further research, especially in inhomogeneous systems.

Another question worthy of further attention is the effect of the collision velocity on the dynamical transition.
We have found more inhomogeneous configurations when $H$ in (\ref{eq:disturbance3}) is much larger than the critical value $H_{*}$.
We leave these questions for further study in the future.

\section*{Acknowledgement}
Q. Chen would like to thank Fenfen Luo for her supporting his work. Y. Liu is grateful to Cheng Peng and Jia Tian for their helpful discussions. He also thanks his wife for her supporting his work. This work is supported in part by the National Natural Science Foundation of China with Grant Nos. 11975235, 12005077, 12035016 and 12075026, as well as by China Postdoctoral Science Foundation, under the National Postdoctoral Program for Innovative Talents BX2021303 and Guangdong Basic and Applied Basic Research Foundation under Grant No. 2021A1515012374. B. W. was partially supported by NNSFC under grant No.12075202. 
%=======================================================================
\bibliographystyle{unsrt}

\bibliography{references}
\end{document}